\documentclass[aps,
reprint,
twocolumn,
prl,
superscriptaddress,
nofootinbib,
]{revtex4-2}
\usepackage[utf8]{inputenc}
\usepackage{amsmath}
\usepackage{amssymb}
\usepackage{anyfontsize}
\usepackage{amsfonts,amsthm,bm}
\usepackage{color,xcolor}
\usepackage{comment}
\usepackage{soul}
\usepackage{chngcntr}
\counterwithout{equation}{subsection}
\counterwithout{equation}{section}

\usepackage[
breaklinks=true,
colorlinks=true,
bookmarksopen,
bookmarksnumbered,
linkcolor=teal,
filecolor=DarkBlue,
urlcolor=PRDblue,
citecolor=tclr,
pdftex
]{hyperref}

\usepackage{graphicx,epsfig}
\usepackage{float}
\usepackage{subcaption}
\usepackage{tikz}

\definecolor{azure(colorwheel)}{rgb}{0.0, 0.5, 1.0}
\definecolor{DarkViolet}{RGB}{148,0,211}
\definecolor{myDarkBlue}{rgb}{0,0.1,0.7}
\definecolor{tealgreen}{rgb}{0.0, 0.51, 0.5}
\definecolor{PRDblue}{RGB}{48,46,146}

\definecolor{tclr}{RGB}{148,0,211}
\newcommand{\SectionColor}[1]{\textcolor{PRDblue}{#1}}
\newcommand{\SectionStyle}[1]{\SectionColor{\textit{\textbf{#1.\hspace{0em}}}}}

\newcommand*\dd{\mathop{}\!\mathrm{d}}
\usepackage{hyperref}
\hypersetup{colorlinks,bookmarksopen,
	bookmarksnumbered,
	citecolor={DarkViolet},
	linkcolor={myDarkBlue},
	urlcolor={tealgreen},
	pdfstartview=FitH}

\usepackage{times}

\usepackage[top=1in, bottom=0.4in, left=0.55in, right=0.55in,includefoot]{geometry}

\graphicspath{{Figs/}}

\usepackage{scalerel}
\usetikzlibrary{svg.path}
\definecolor{orcidlogocol}{HTML}{A6CE39}
\tikzset{orcidlogo/.pic={\fill[orcidlogocol] svg{M256,128c0,70.7-57.3,128-28,128C57.3,256,0,198.7,0,128C0,57.3,57.3,0,128,0C198.7,0,256,57.3,256,128z};\fill[white]svg{M86.3,186.2H70.9V79.1h15.4v48.4V186.2z}svg{M108.9,79.1h41.6c39.6,0,57,28.3,57,53.6c0,27.5-21.5,53.6-56.8,53.6h-41.8V79.1z M124.3,172.4h24.5c34.9,0,42.9-26.5,42.9-39.7c0-21.5-13.7-39.7-43.7-39.7h-23.7V172.4z}svg{M88.7,56.8c0,5.5-4.5,10.1-10.1,10.1c-5.6,0-10.1-4.6-10.1-10.1c0-5.6,4.5-10.1,10.1-10.1C84.2,46.7,88.7,51.3,88.7,56.8z};}}
\newcommand\orcid[1]{\href{https://orcid.org/#1}{\mbox{\scalerel*{
\begin{tikzpicture}[yscale=-1,transform shape]
\pic{orcidlogo};
\end{tikzpicture}
}{|}}}}

\usepackage{orcidlink}
\def\idmyung{\orcidlink{0000-0001-6343-9959}}
\def\idnakas{\orcidlink{0000-0002-3522-5803}}

\begin{document}

\title{\Large Smarr formula for black holes with primary and secondary scalar hair}

\author{Yun Soo Myung\,\idmyung}
\email{ysmyung@inje.ac.kr}
\affiliation{Center for Quantum Spacetime, Sogang University, Seoul 04107,  Korea}

\author{Theodoros Nakas\,\idnakas}
\email{theodoros.nakas@gmail.com}
\affiliation{Cosmology, Gravity, and Astroparticle Physics Group, Center for Theoretical Physics of the Universe, Institute for Basic Science (IBS), Daejeon 34126, Korea}

 \begin{abstract}
    \noindent\textbf{Abstract.}
	In this work, we revisit the thermodynamics of black holes endowed with primary and secondary scalar hair in the shift and ${\rm Z}_2$ symmetric subclass of beyond Horndeski gravity.
    Under a specific fine-tuning of the scalar parameter $q$ in terms of the black hole mass, the singular black-hole solution with primary scalar hair reduces to the regular Bardeen solution featuring secondary scalar hair. 
    We first demonstrate that the traditional thermodynamic approach fails to yield a consistent Smarr formula for both solutions under consideration. 
    To address this issue, we adopt the approach introduced in [Phys.Rev.Lett.\,132\,(2024)\,19,\,191401], and we derive both the first law of black hole thermodynamics and the Smarr formula, offering a consistent thermodynamic description for scalar-hairy black holes.
    As an additional outcome, our analysis reveals a connection between the solutions with primary and secondary scalar hair.
\end{abstract}

\maketitle

\flushbottom


\vspace{2mm}


\SectionStyle{Introduction} No-scalar-hair theorems (see e.g. \cite{Bekenstein:1996pn, Herdeiro:2015waa} and references therein) once suggested that black holes are fully characterized by mass, angular momentum, and charge (if present).
However, the last two decades, a growing body of work has demonstrated that black holes can indeed acquire scalar hair in modified theories of gravity \cite{Herdeiro:2015waa, Barack:2018yly}. 
Among these, Horndeski \cite{Horndeski:1974wa} and beyond Horndeski gravity \cite{Zumalacarregui:2013pma, Gleyzes:2014dya, Langlois:2015cwa, Langlois:2018dxi} provide a fertile ground for exploring such deviations where solutions with primary or secondary scalar hair emerge naturally, and particularly in theories exhibiting shift and ${\rm Z}_2$ symmetries with respect to the scalar field \cite{Babichev:2013cya, Bakopoulos:2022csr, Bakopoulos:2023fmv, Bakopoulos:2023sdm}.
Solutions with primary scalar hair feature, in addition to the aforementioned standard parameters, an extra independent quantity directly associated with the scalar field.
In contrast, solutions with secondary scalar hair, although also arising from scalar-tensor theories, are entirely characterized by the mass of the black hole.  

These developments naturally raise the question of how such scalar-hairy black holes behave thermodynamically, and whether the established thermodynamic framework, including the Smarr formula, can be consistently extended to accommodate them.
The study of black holes from a thermodynamic perspective plays a central role in gravitational physics, offering new insights for our deeper understanding of gravity \cite{Bekenstein:1973ur, Bardeen:1973gs, Hawking:1975vcx, Carlip:2014pma, Padmanabhan:2009vy, Susskind:1993if}. 
Although the thermodynamic analysis of solutions within General Relativity (GR) is well established \cite{Bekenstein:1973ur, Bardeen:1973gs, Hawking:1975vcx}, extending the same analysis to solutions originating from modified gravitational theories is not always straightforward and, in some cases, fails to yield meaningful results.
It is often the case that new insights and interpretations are needed in order to accommodate the new features of black holes.
Especially for regular black holes, i.e. compact objects that possess event horizons but are devoid of inner singularities, the first law of black hole thermodynamics requires modification as it is discussed in \cite{Ma:2014qma}.
Another interesting aspect of GR black holes, which is often difficult to be satisfied for black-hole solutions coming from scalar-tensor theories, is that their mass always admits a particular decomposition which involves both thermodynamic quantities, such as the temperature and the entropy, as well as other physical quantities, such as the angular velocity and the electric potential.
This decomposition is known as Smarr formula \cite{Smarr:1972kt}, named after Larry Smarr, who first derived it for the most general GR black hole, that is the Kerr-Newman solution.

In an attempt to address the aforementioned inconsistencies arising in beyond-GR black holes, various approaches have been developed over the years.
One compelling new approach was recently proposed in \cite{Hajian:2023bhq}, where the dimensionful coupling constants---typically fixed in the theory---are promoted to conserved charges associated with auxiliary fields. 
In this formulation, the auxiliary fields are introduced in such a way that the underlying solution remains unchanged.
A direct consequence of this approach is that any coupling constant previously appearing in the solution as a fixed constant, should now be considered a free parameter and treated on equal footing with the other thermodynamic variables.

In this work, we adopt the approach of \cite{Hajian:2023bhq} to study black holes with primary and secondary scalar hair, treating each case separately. 
These solutions arise within the shift and ${\rm Z}_2$ symmetric subclass of beyond Horndeski gravity, as presented in \cite{Bakopoulos:2023sdm}, and are connected through a specific fine-tuning of the scalar hair parameter $q$ in terms of the black-hole mass.
In the case of secondary scalar hair, the solution that emerges is the well-known Bardeen solution, which in this setup arises as a solution of the beyond Horndeski theory rather than from a non-linear electromagnetic theory with magnetic monopole configuration \cite{Ayon-Beato:2000mjt}.
We derive both the first law of black hole thermodynamics and the Smarr formula for these two classes of solutions and show that the black-hole mass plays the role of the internal energy of the system in every scenario. 
Additionally, the connection between the solutions with primary and secondary scalar hair becomes apparent both at the level of the first law and in the structure of the Smarr formula.


\SectionStyle{Black holes with primary and secondary scalar hair in beyond Horndeski gravity} We consider the shift ($\phi \rightarrow \phi+ {\rm const.}$) and ${\rm Z}_2$ ($\phi\rightarrow -\phi$) symmetric beyond Horndeski gravity in geometrized units ($c=G=1$) given by the following action functional
\begin{equation}
\label{eq:act}
    \mathcal{S}=\frac{1}{16\pi} \int \dd^4x\sqrt{-g}\big(\mathcal{L}^H_2+\mathcal{L}^H_4+\mathcal{L}^{bH}\big)\,,
\end{equation}
where
\begin{align}
    &\mathcal{L}^{H}_2 = G_2(X)\,,\\[2mm]
    &\mathcal{L}^{H}_4 = G_4(X)R + G_{4X}\left[ (\square\phi)^2-(\nabla_\mu\nabla_\nu\phi) (\nabla^\mu\nabla^\nu\phi) \right]\,,\\[2mm]
    &\mathcal{L}^{bH}=F_4(X)\,\epsilon^{\mu\nu\rho\sigma}\epsilon^{\alpha\beta\gamma}{}_\sigma\, \partial_\mu\phi\, \partial_\alpha\phi\, (\nabla_\nu \nabla_\beta \phi)\, (\nabla_\rho\nabla_\gamma \phi)\,.
\end{align}
In the above, $\mathcal{L}_2^H$ and $\mathcal{L}_4^H$ belong to Horndeski theory, while $\mathcal{L}^{bH}$ is the ${\rm Z}_2$-symmetric beyond Horndeski term. 
The latter can emerge from the Horndeski gravity through a disformal transformation of the metric \cite{Zumalacarregui:2013pma, Gleyzes:2014dya, Langlois:2015cwa, Langlois:2018dxi}. 
Note that theory \eqref{eq:act} leads to second order equations of motion without the need of any degeneracy conditions. 
The model functions $G_2$, $G_4$, and $F_4$ are arbitrary functions of the scalar field's kinetic term $X=-\partial^\mu\phi\,\partial_\mu\phi/2$, with dimensions $(\text{length})^2$, $0$ (dimensionless), and $(\text{length})^4$.
Throughout this article, we follow the notation that a subscript $X$ in the model functions denotes derivative with respect to $X$, namely $G_{iX}=\dd G_i/\dd X$.

The shift symmetric nature of the theory \eqref{eq:act} allows, in addition to radial, a linear time dependence on the scalar field
\begin{equation}
    \label{eq: scalar}
    \phi=qt+\psi(r)\,,
\end{equation}
whereas the staticity of the metric remains unaffected \cite{Babichev:2013cya, Kobayashi:2014eva, Babichev:2017guv}.
It was recently realized \cite{Bakopoulos:2023fmv, Bakopoulos:2023sdm} that the action \eqref{eq:act} gives rise to homogeneous black-hole solutions endowed with primary scalar hair and charge.
Specifically, for 
\begin{equation}
	\label{eq:model-fun}
    \begin{gathered}
    G_2(X)=-\frac{2\eta}{\lambda^2}X^{5/2}\,,\quad G_4(X)=1-\eta X^{5/2}\,,\\[1mm]
    F_4(X)=\eta \sqrt{X}\,,
    \end{gathered}
\end{equation}
where $\eta$ and $\lambda$ are coupling constants, with dimensions $(\text{length})^{5}$ and $(\text{length})$, respectively, one finds the solution
\begin{equation}
    \label{eq:sol}
    \begin{gathered}
    \dd s^2=-h(r)\dd t^2+\frac{\dd r^2}{h(r)}+r^2\dd\Omega^2\,,\\[1mm]
    h(r)=1-\frac{2M}{r}+\frac{\sqrt{2}\,\eta q^5}{3}\frac{\lambda}{r}\bigg(1-\frac{r^3}{(r^2+\lambda^2)^{3/2}} \bigg)\,.
    \end{gathered}
\end{equation}
In the above, $\dd \Omega^2\equiv \dd\vartheta^2+\sin^2\vartheta \dd\varphi^2$ represents the line element of the unit sphere, the parameter $M$ describes the Arnowitt-Deser-Misner (ADM) mass \cite{Arnowitt:1961zz}, while $q$ is the primary scalar hair of the solution and is completely independent of the mass. 
This distinguishes it from cases where there is an explicit dependence between $q$ and $M$, scenarios that result in solutions with secondary scalar hair. 
In the upcoming analysis, we consider that the parameters $\eta$, $\lambda$, and $q$ take positive values, for which one can show that the line element \eqref{eq:sol} describes black holes with one or more horizons.
Note also, that in the class of theories where the model function $G_4(X)$ is a linear function of $G_2(X)$, which is the case under consideration here, the kinetic term of the scalar field is always given by the following relation
\begin{equation}
X=\frac{q^2}{2}\frac{1}{1+(r/\lambda)^2}\,.
\end{equation}

The internal shift symmetry of the theory, generates the Noether current
\begin{equation}
	\label{eq:curr}
	\mathbf{J}=J_\mu \dd x^\mu = \frac{1}{\sqrt{-g}} \frac{\delta \mathcal{S}}{\delta (\partial_\mu \phi)}\,, \hspace{1em}J^\mu=-\frac{4X}{q}\, G_{2X}\, \delta^\mu_0\,.
\end{equation}
The charge stemming from the Noether current \eqref{eq:curr}, evaluated on a spacelike hypersurface, is associated with the scalar hair $q$, as it can be easily confirmed through a direct computation
\begin{equation}
	\label{eq:Qs}
	Q_s \propto \int \star \mathbf{J} \propto \frac{20\pi}{3\sqrt{2}}\, \lambda \eta q^4\,.
\end{equation}
Interestingly, by fixing the primary scalar hair $q$ in terms of the mass parameter $M$ as $q^5=\frac{3\sqrt{2}M}{\eta \lambda}$, the $1/r$ term in the solution \eqref{eq:sol} vanishes altogether.
The resulting solution is the Bardeen solution which describes regular (i.e. non-singular) black holes and is given by the line element
\begin{equation}
	\label{eq:bardeen-metr}
	\begin{gathered}
	\dd s^2=-\bar{h}(r)\dd t^2+\frac{\dd r^2}{\bar{h}(r)}+r^2\dd\Omega^2\,,\\[1mm]
	\bar{h}(r)=1-\frac{2Mr^2}{(r^2+\lambda^2)^{3/2}}\,.
	\end{gathered}
\end{equation}
It is important to notice here, that the mass $M$ remains a free parameter, however, due to the tuning of the scalar hair $q$ in terms of the mass $M$, the solution \eqref{eq:bardeen-metr} is of secondary scalar hair.
This can also be deduced from the expression of the conserved scalar charge $\bar{Q}_s$, which in this case takes the form
\begin{equation}
	\label{eq:sec-hair}
	\bar{Q}_s \propto 20\pi \frac{M}{q}=20\pi M \left(\frac{\eta \lambda}{3\sqrt{2}M} \right)^{1/5}\,.
\end{equation} 
We notice that the charge $\bar{Q}_s$ depends solely on the mass of the black hole, which is the only free parameter.
Therefore, the Bardeen solution originating from the beyond Horndeski gravity \eqref{eq:act} and \eqref{eq:model-fun} possesses secondary scalar hair.

Traditionally, the Bardeen solution is interpreted as a solution to a non-linear electromagnetic (NLED) theory which is sourced by a magnetic monopole (see e.g. \cite{Ayon-Beato:2000mjt}) and is described by the following action functional:
\begin{equation}
	\label{eq:bardeen-magn-act}
	\mathcal{S}_{\rm NLED}=\frac{1}{16\pi}\int d^4x\bigg[R-\frac{12M}{|\lambda|\lambda^2}\bigg(\frac{\sqrt{2\lambda^2 F}}{1+\sqrt{2\lambda^2 F}} \bigg)^{5/2}\, \bigg]\,.
\end{equation}
In the above, $F\equiv F^{\mu\nu}F_{\mu\nu}/4$ is the electromagnetic field strength invariant, the components of the electromagnetic tensor are evaluated via $F_{\mu\nu}=2\lambda \delta^\vartheta_{[\mu} \delta^\varphi_{\nu]}\sin\vartheta$, while, in this case, the parameter $\lambda$ is the magnetic monopole charge.
In contrast to the Bardeen solution derived from the beyond Horndeski theory, where the mass parameter $M$ of the solution emerges independently of the theory’s couplings, we see that in this case, both the mass $M$ and the magnetic  charge $\lambda$ explicitly appear as coupling constants in the action \eqref{eq:bardeen-magn-act}. 
As a result, specifying the theory by assigning values to its coupling constants simultaneously determines the parameters of the corresponding solution \eqref{eq:bardeen-metr}.
Consequently, although \eqref{eq:bardeen-magn-act} is commonly regarded in the literature as the theory that gives rise to the Bardeen solution, we emphasize here that the beyond Horndeski gravity admits a broader family of regular black-hole solutions with different masses.


\SectionStyle{Standard black hole thermodynamics} The thermodynamic analysis for the black-hole solutions originating from the beyond Horndeski theory \eqref{eq:act} has been performed in \cite{Bakopoulos:2024ogt} following the Grand Canonical Ensemble (GCE) approach, and using the Euclidean action.
Here, upon adopting the GCE approach and using the differential form for the black-hole mass \cite{PhysRevD.4.3552, PhysRevLett.25.1596, Smarr:1972kt}, we verify the results presented in \cite{Bakopoulos:2024ogt} for both \eqref{eq:sol} and \eqref{eq:bardeen-metr} solutions. 
We demonstrate that, for the black-hole solutions with primary scalar hair \eqref{eq:sol}, the black-hole mass can be interpreted as the internal energy of the system corresponding to the sum of the \textit{surface energy} and the \textit{scalar energy} of the black hole. 
In contrast, we find that the first law of black hole thermodynamics is not satisfied for the Bardeen solution with secondary hair, unless additional assumptions are imposed. 
Furthermore, we show that the Smarr relation cannot be consistently satisfied for either of these two classes of solutions.
A solution to the aforementioned inconsistencies will be presented in the next section.
In this section, our goal is to showcase how these problems emerge. 


Starting from solution \eqref{eq:sol}, and solving the equation $h(r_h)=0$, where $r_h$ represents the event horizon of the black hole,\,\footnote{The black-hole horizon $r_h$ is defined as the largest positive root to the equation $h(r_h)=0$.} with respect to the ADM mass $M$, and using also the fact that the black-hole entropy is given by the usual area law $S=A/4=\pi r_h^2$, one finds that
\begin{equation}
    \label{eq:mass}
    M = \frac{\sqrt{S}}{2\sqrt{\pi }}+\frac{\eta q^5 \lambda}{3\sqrt{2}}  \left[1-\frac{S^{3/2}}{\left(S+\pi  \lambda ^2\right)^{3/2}}\right]\,.
\end{equation}
Using \eqref{eq:mass}, and considering the entropy $S$ and the scalar hair $q$ as the independent parameters, one can readily obtain the first law of black hole thermodynamics in its differential form, namely
\begin{equation}
	\label{eq:1st-law1}
	\dd M=T_H \dd S+\Phi_q \dd q\,.
\end{equation}
The temperature $T_H$ of the black hole, given by
\begin{equation}
	\label{eq:temp}
	T_H\equiv \frac{\partial M}{\partial S}=\frac{1}{4\sqrt{\pi}\sqrt{S}}-
	\frac{\pi  \lambda ^3 \eta  q^5 \sqrt{S}}{2\sqrt{2}\left(S+\pi  \lambda ^2\right)^{5/2}}\,,
\end{equation}
is equivalent to $T_h\equiv \frac{1}{4\pi}\frac{\dd h}{\dd r}\big|_{r=r_h}$, as one can verify with a straightforward computation.
The quantity $\Phi_q$ can be interpreted as the chemical potential associated with the scalar hair $q$ and is expressed as
\begin{equation}
	\label{eq:chem-q}
	\Phi_q\equiv \frac{\partial M}{\partial q}=\frac{5\lambda \eta q^4}{3 \sqrt{2}}\left[1-\frac{S^{3/2} }{ \left(S+\pi  \lambda ^2\right)^{3/2}}\right]\,.
\end{equation}
Following the analysis in \cite{Smarr:1972kt}, and since $\dd M$ is a perfect differential in \eqref{eq:1st-law1}, one is free to choose any convenient path of integration in the $(S,q)$ space.
By choosing the path in an appropriate way, one is able to define two energy components for the solution \eqref{eq:sol}.
The first one is the surface energy $E_s$ which is given by
\begin{equation}
	\label{eq:Es}
	E_s=\int_0^S T_H(\tilde{S},0) \dd \tilde{S}=\frac{\sqrt{S}}{2\sqrt{\pi}}\,,
\end{equation}
and is common even for the GR black holes, i.e. Schwarzschild, Kerr, and Kerr-Newman, while the second one is the scalar energy $E_q$ related to the scalar hair $q$ and has the form
\begin{equation}
	\label{eq:Eq}
	E_q=\int_0^q \Phi_q(S,\tilde{q}) \dd \tilde{q}=\frac{\eta q^5 \lambda}{3\sqrt{2}}  \left[1-\frac{S^{3/2}}{\left(S+\pi  \lambda^2\right)^{3/2}}\right]\,.
\end{equation} 
We observe that $M=E_s+E_q$ indicating that the mass of the black hole plays the role of the internal energy of the system.
This type of thermodynamic decomposition is also typical in GR black-hole solutions, however, instead of the scalar energy, the rotational energy and/or the electromagnetic energy appear (see e.g. \cite{Smarr:1972kt}).
The key distinction in this case is that the Smarr formula cannot be satisfied, unlike for the GR black-hole solutions.
By using eqs. \eqref{eq:mass}, \eqref{eq:temp}, and \eqref{eq:chem-q}, one can show through a direct computation that  
\begin{equation}
	M\neq c_{s} T_H S+ c_q \Phi_q q\,,
\end{equation}
for any $c_s,\, c_q \in \mathbb{R}$, thereby rendering a Smarr-like decomposition impossible.


For the Bardeen solution \eqref{eq:bardeen-metr}, solving the equation $\bar{h}(r_h)=0$ with respect to the black-hole mass and using the area law $S=\pi r_h^2$ for the entropy, one obtains the relation
\begin{equation}
	\label{eq:bardeen-mass}
	M=\frac{\left(\pi  \lambda ^2+S\right)^{3/2}}{2 \sqrt{\pi } S}\,.
\end{equation}
As in the previous solution, the parameter $\lambda$ is a coupling constant, and therefore the differential form of the mass should be of the form 
\begin{equation}
	\label{eq:bardeen-1st-law1}
	\dd M = T_H \dd S\,,
\end{equation}
with the temperature $T_H$ given by 
\begin{equation}
	\label{eq:bardeen-temp}
	T_H=\sqrt{S+\pi\lambda ^2}\, \bigg(\frac{1}{4\sqrt{\pi}\, S}-\frac{\sqrt{\pi}\,  \lambda ^2}{2 S^2} \bigg)\,.
\end{equation}
Here, one can verify that $T_H\neq T_h\equiv \frac{1}{4\pi}\frac{\dd \bar{h}}{\dd r}\big|_{r=r_h}$ and also the surface energy $E_s$, when evaluated in the same way as in the previous example, diverges.
As a result, eq.\,\eqref{eq:bardeen-1st-law1} fails to describe correctly the first law of black hole thermodynamics, and consequently, a Smarr-like decomposition is also impossible in this case.
In \cite{Ma:2014qma}, the authors propose a method to consistently derive the first thermodynamic law for regular black holes by employing the stress-energy tensor.
In the following section, we adopt a different approach, nevertheless, the first law for the Bardeen solution will be proven to be equivalent with the result presented in \cite{Ma:2014qma}.


\SectionStyle{Black hole thermodynamics and Smarr formula revisited} In this section, we reexamine the thermodynamics of the preceding black-hole solutions through the lens of a recently proposed interpretation introduced in \cite{Hajian:2023bhq}. 
The core idea of this new framework is to incorporate, in addition to the usual free parameters, all dimensionful coupling constants appearing in the solution into the thermodynamic analysis. 
This can be achieved due to the fact that for any covariant Lagrangian, the coupling constants can be consistently promoted to conserved charges \cite{Hajian:2023bhq}.
The above procedure is made possible by introducing auxiliary fields in the original theory, in such a way that the background solution remains completely unchanged.
As a result, the previously fixed coupling constants are now appearing into the solution as free parameters, and for that reason, they should be considered on equal footing with any other thermodynamic variable.


Our starting point is the same as in previous analysis, that is the line element \eqref{eq:sol} and the expression for the mass in terms of the thermodynamic variables evaluated on the black-hole horizon, eq.\,\eqref{eq:mass}.
The key difference now is that the previously fixed dimensionful coupling constants, $\eta$ and $\lambda$, will be treated as free parameters, in accordance with the preceding discussion and the framework introduced in \cite{Hajian:2023bhq}.
Consequently, the first thermodynamic law that we need to consider is of the form
\begin{equation}
	\label{eq:1st-law2}
	\dd M= T_H \dd S+ \Phi_q \dd q+ \Phi_\eta \dd \eta + \Phi_\lambda \dd \lambda\,,
\end{equation}
where the expressions for both the temperature $T_H$ and the chemical potential $\Phi_q$ remain the same as before, namely eqs.~\eqref{eq:temp}, \eqref{eq:chem-q}.
In addition to these two thermodynamic quantities, the chemical potentials $\Phi_\eta\equiv \partial M/\partial \eta$ and $\Phi_\lambda\equiv\partial M/\partial \lambda$ should also be taken into account, for which we find the expressions
\begin{align}
	\label{eq:chem-eta}
	&\Phi_\eta=-\frac{\lambda  q^5}{3 \sqrt{2}}\left[1-\frac{S^{3/2} }{ \left(S+\pi  \lambda ^2\right)^{3/2}}\right]\,,\\[2mm]
	\label{eq:chem-lam}	
	&\Phi_\lambda=-\frac{q^5 \eta  \left[S^{5/2}-2 \pi  \lambda ^2 S^{3/2}-\left(S+\pi  \lambda ^2\right)^{5/2}\right]}{3 \sqrt{2}
	\left(S+		\pi  \lambda ^2\right)^{5/2}}\,.
\end{align}
It is of crucial importance to notice at this point that the combination $\eta q^5$ of the free parameters $\eta$ and $q$ is dimensionless.
Consequently, we can define the positive\,\footnote{We remind the reader that $\eta$, $\lambda$, and $q$ are assumed to take positive values.} dimensionless parameter $\xi$ in the following way
\begin{equation}
	\label{eq:xi-def}
	\xi\equiv \eta q^5\,, \hspace{1em} [\xi]=0 \text{ (dimensionless)}\,.
\end{equation}
By doing so, it is straightforward to deduce that  $\dd\eta = -5\xi q^{-6} \dd q$.
Combining now the previous relation with eqs.~\eqref{eq:chem-q} and \eqref{eq:chem-eta} we find that $\Phi_q \dd q+\Phi_\eta \dd\eta$ vanishes identically indicating that parameter $\eta$ is not independent of $q$.
Hence, the first thermodynamic law of eq.~\eqref{eq:1st-law2} reduces to
\begin{equation}
	\label{eq:1st-law3}
	\dd M= T_H \dd S + \Phi_\lambda \dd \lambda\,,
\end{equation}
In fact, the above result can be directly obtained from the line element \eqref{eq:sol} by making use of \eqref{eq:xi-def} and solving $h(r_h)=0$ in terms of the mass and then evaluating the differential $\dd M$ with respect to the free dimensionful parameters $S$ and $\lambda$.

Due to the fact that $\dd M$ is a perfect differential, one is free to choose any convenient path of integration in the $(S,\lambda)$ space.
In this way, each of the free parameters corresponds to a distinct energy component.
The path that we will follow is $(0,0)\rightarrow (S,0) \rightarrow (S,\lambda)$.
Hence, for the surface energy $E_s$ we have
\begin{equation}
	\label{eq:Es-new}
	E_s=\int_0^S T_H(\tilde{S},0) \dd \tilde{S}=\frac{\sqrt{S}}{2\sqrt{\pi}}\,,
\end{equation}
which is the same as before, while for the energy corresponding to the $\lambda$-parameter one finds
\begin{equation}
	\label{eq:E-lam}
	E_\lambda=\int_0^\lambda \Phi_\lambda(S,\tilde{\lambda}) \dd \tilde{\lambda}=\frac{\xi \lambda}{3\sqrt{2}}  
	\left[1-\frac{S^{3/2}}{\left(S+\pi \lambda^2\right)^{3/2}}\right]\,.
\end{equation}
We notice that 
\begin{equation}
	\label{eq:mass-int-ene}
	M=E_s+E_\lambda\,,
\end{equation}
which means that the black-hole mass plays the role of the internal energy of the system, as expected.
Moreover, a direct computation confirms that the Smarr decomposition holds in this case satisfying the form
\begin{equation}
	\label{eq:smarr-prim}
	M=2 T_H S+ \Phi_\lambda \lambda\,.
\end{equation}
Note also that upon using relation \eqref{eq:xi-def}, the expression for $E_\lambda$ coincides with that of $E_q$ in eq.~\eqref{eq:Eq}.
This leads us to consider the connection between the parameter $\lambda$ and the primary scalar hair of the black hole \eqref{eq:sol}. 
Equation \eqref{eq:Qs} dictates that by choosing an appropriate proportionality constant, multiplying the nominator and denominator of the r.h.s. with $q$, and using \eqref{eq:xi-def}, one obtains $\lambda=Q_s q$.
Consequently, eq.~\eqref{eq:1st-law3} can be expressed as
\begin{equation}
	\label{eq:1st-law-4}
	\dd M=T_H \dd S+\Phi_q \dd q+\Phi_{Q_s}\dd Q_s \,,
\end{equation}
where $\Phi_{Q_s}=q\,\Phi_\lambda$ and $\Phi_q=Q_s\Phi_\lambda$.
Employing the same method and using the integration path $(0,0,0)\rightarrow (S,0,0)\rightarrow (S,Q_s,0)\rightarrow (S,Q_s,q)$, it is straightforward to show that
\begin{equation}
	\label{eq:mass-int-ene-new}
	M=E_s+E_{q}
\end{equation}
since
{\fontsize{8pt}{12pt}\selectfont
\begin{align}
	\label{eq:Es-new-2}
	&E_s=\int_0^S T_H(\tilde{S},0,0) \dd \tilde{S}=\frac{\sqrt{S}}{2\sqrt{\pi}}\,,\\[2mm]
	&E_{Q_s}=\int_0^{Q_s} \Phi_{Q_s}(S,\tilde{Q}_s,0) \dd \tilde{Q}_s=0\,,\\[2mm]
	\label{eq:Eq-new}
	&E_{q}=\int_0^{q} \Phi_{q}(S,Q_s,\tilde{q}) \dd \tilde{q}=\frac{\xi q Q_s}{3\sqrt{2}}  
	\left[1-\frac{S^{3/2}}{\left(S+\pi \lambda^2\right)^{3/2}}\right]\,\,.
\end{align}}
As for the Smarr decomposition, one can readily show that
\begin{equation}
	M=2T_H S+c_{Q_s}\Phi_{Q_s} Q_s+c_{q}\Phi_{q} q\,,
\end{equation} 
for all $c_q$, $c_{Q_s}$ that satisfy $c_q+c_{Q_s}=1$.
We see that eq.~\eqref{eq:mass-int-ene-new} matches the form of the standard thermodynamic analysis, while the expressions for $E_s$ and $E_q$ are given by the same expressions upon using the relations $\xi=\eta q^5$ and $\lambda=q Q_s$.  
Note that $E_{Q_s}=0$ and $c_q+c_{Q_s}=1$ simply reflect the fact that the parameters $Q_s$ and $q$ are not independent of each other, but rather interconnected through $\lambda$. 
In fact, depending on the path of integration, one can make $E_q$ to vanish instead of $E_{Q_s}$ by maintaining the same result.

From the above analysis, we deduce that upon promoting the dimensionful coupling constants $\eta$ and $\lambda$ to free parameters exploiting the results of \cite{Hajian:2023bhq}, makes it possible to satisfy the Smarr formula for the black holes with primary scalar hair described by eq.~\eqref{eq:sol}.
Interestingly, this is achieved without spoiling the interpretation of the black-hole mass as the internal energy of the system.  
Finally, performing an appropriate reformulation of the first thermodynamic law, the black-hole mass is expressed as the sum of the black hole's surface energy and the energy $E_{q}$ (or $E_{Q_s}$), which is associated with the existence of the primary scalar charge.


Pertaining now to the Bardeen solution \eqref{eq:bardeen-metr} and incorporating the parameter $\lambda$ in the first thermodynamic law \cite{Hajian:2023bhq}, eq.~\eqref{eq:bardeen-mass} lead us to
\begin{equation}
	\label{eq:bardeen-1st-law}	
	\dd M= T_H \dd S+\Phi_\lambda \dd\lambda\,,
\end{equation} 
where $T_H$ is given by \eqref{eq:bardeen-temp} and the chemical potential $\Phi_\lambda$ has the form
\begin{equation}
	\label{eq:bard-chem}
	\Phi_\lambda=\frac{3 \sqrt{\pi }\, \lambda  \sqrt{S+\pi  \lambda ^2}}{2 S}\,.
\end{equation}
Following the $(0,0)\rightarrow (S,0) \rightarrow (S,\lambda)$ path of integration in this case, it is straightforward to evaluate the surface energy and the energy associated with the $\lambda$ parameter.
By doing so, we are led to
\begin{align}
	\label{eq:bard-Es}
	&E_s=\int_0^S T_H(\tilde{S},0) \dd \tilde{S}=\frac{\sqrt{S}}{2\sqrt{\pi}}\,,\\[2mm]
	\label{eq:bard-E-lam}
	&E_\lambda=\int_0^\lambda \Phi_\lambda(S,\tilde{\lambda}) \dd \tilde{\lambda}=\frac{\left(S+\pi  \lambda ^2\right)^{3/2}-S^{3/2}}{2 \sqrt{\pi } S}\,.
\end{align}
As expected, we observe that $M=E_s+E_\lambda$, indicating that the mass can be interpreted as the internal energy of the system, something that was not possible without treating the parameter $\lambda$ as a free thermodynamic variable.
The above analysis of the first thermodynamic law is in agreement with the result obtained in \cite{Ma:2014qma}, where the authors employed a completely different approach.

As for the Smarr decomposition, it is straightforward to verify that this holds for
\begin{equation}
	\label{eq:bardeen-smarr}
	M=2T_H S+\Phi_\lambda \lambda\,.
\end{equation}
It is interesting to notice, that both the first thermodynamic law and the Smarr formula for the solutions with primary and secondary scalar hair have exactly the same form.
This suggests that both solutions have a shared underlying origin, which is in fact the case.


\SectionStyle{Conclusions} In this work, we revisited the thermodynamics of black holes possessing both primary and secondary scalar hair within the shift and ${\rm Z}_2$ symmetric subclass of beyond Horndeski gravity \cite{Bakopoulos:2023sdm}.
The solution with secondary scalar hair is the Bardeen solution which appears a special case of the singular solution with primary scalar hair under a particular fine-tuning of the scalar parameter $q$ in terms of the black-hole mass.  
Our analysis had two objectives: first, to derive the first law of black hole thermodynamics for these solutions, and second, to investigate whether a Smarr formula can be established despite the presence of scalar hair.
Initially, we employed the standard thermodynamic approach and demonstrated its limitations.
Specifically, for the Smarr decomposition, the standard approach was proven to be unable to yield such an expression for the black holes considered.

To overcome this limitation, we adopted the recent approach introduced in \cite{Hajian:2023bhq}, which promotes, in a systematic way, the dimensionful coupling constants of any covariant Lagrangian to conserved charges through the inclusion of auxiliary fields. 
Despite the addition of auxiliary fields, this method preserves the original solution and allows the new charges to participate in the thermodynamic description. 
While unconventional at first glance, this approach has proven highly effective in describing simultaneously both the first thermodynamic law and the Smarr formula of solutions in modified gravitational theories. 
By applying this approach to the solutions under consideration, we successfully derived the first law and established the corresponding Smarr formula, thereby highlighting the utility of this extended thermodynamic framework for black holes with scalar hair.
As an additional byproduct of our analysis, we observed a connection between solutions with primary and secondary scalar hair, evident both in the form of the first thermodynamic law and in the structure of the Smarr formula.


\SectionStyle{Acknowledgments} Y.S.M. was supported by the National Research Foundation of Korea(NRF) grant
 funded by the Korea government(MSIT) (RS-2022-NR069013). T.N.  is supported by IBS under the project code IBS-R018-D3.

\bibliographystyle{apsrev4-1}
\bibliography{Refs.bib}

\end{document}